# Coupling of electron spin with its rotation in semiconductors.

Yuri A. Serebrennikov

Qubit Technology Center

2152 Merokee Dr., Merrick, NY 11566

The interplay between spin-orbit interaction in semiconductor valence bands and an adiabatic rotational distortion of the wave function of a charge carrier leads to the scalar spin-orbit-rotation term in the effective-mass Hamiltonian of the conduction-band electron. The physical origin of this result lies in the fact that, similarly to magnetic field effects, the motion of a particle and the phase of its wave function may be affected by the vector potential of the inertial Coriolis field. Here we present a straightforward derivation of this interaction within the multiband envelope function approximation.

72.80.Ey 72.25.Dc 72.25.Hg

In general, the coupling of the total angular momentum of a particle with its rotation yields "forces" of rotational inertia that work on a quantum level[1]. It is well known that the quantum mechanical description of the motion of a spinless particle in the non-inertial rotating ($R$) frame to first order in the angular velocity $\omega$ is formally identical to the account of its motion in the presence of a weak magnetic field[2,3]. To see this one should replace the vector potential of a magnetic field $\vec{A} = \vec{B} \times \vec{r}/2$ with $(2m_0 c/q)\vec{A}_\omega$, where $\vec{A}_\omega = \vec{\omega} \times \vec{r}/2$ is the vector potential of the Coriolis field[4] in the corresponding wave equation ($c$ is the speed of light, $q$ is the charge, and $m_0$ is the mass of a particle). This is also apparent if we compare the expression for the canonical momentum of a charged particle in the presence of magnetic field $\vec{p} = m_0\vec{v} + (q/c)\vec{A}$ with the corresponding expression in the *R*-frame $\vec{p} = m_0\vec{v} + 2m_0\vec{A}_\omega$ at zero magnetic fields, $B = 0$.

For spin bearing particles, the total angular momentum is the sum of the orbital and spin contributions. The non-relativistic *R*-frame Hamiltonian of a *free* spin-1/2 particle at zero magnetic fields can be written as[5]

$$H_\omega^{(R)} = p^2/2m_0 - \hbar\vec{\omega}\cdot\vec{j}, \qquad (1)$$

where $\vec{p}$ is the canonical momentum, $\vec{j} = \vec{l} + \vec{\sigma}/2$ is the total angular momentum, $\hbar\vec{l} = \vec{r}\times\vec{p}$ is the orbital momentum, and $\vec{\sigma}$ is the 3-vector of Pauli matrices. Notably this Hamiltonian incorporates the Mashhoon spin-rotation interaction, $-\hbar\vec{\omega}\cdot\vec{S}$, even in the absence of relativistic spin-orbit (SO) coupling. A strong electric field near heavy nuclei (Ge, Ga, In, etc.) in common semiconductors leads, however, to a strong *intrinsic* SO interaction in the valence bands. The interplay between this interaction and an adiabatic rotational distortion of the wave function of a charge carrier yields the scalar spin-orbit-rotation (SOR) coupling term in the effective-mass Hamiltonian of the conduction-band electron. In our recent article[6] we demonstrated that the SOR coupling can be described in purely geometric terms as a consequence of the difference in the Berry phase acquired by the components of the spin-orbitally mixed Kramers-doublet during its cyclic evolution in the reciprocal momentum space. The physical origin of this result lies in the fact that, similarly to magnetic field effects, the motion of a particle and the phase of its wave function may be affected by the vector potential of the inertial Coriolis field. Here we present a straightforward derivation of the SOR interaction within the multiband envelope function (EF) approximation[7].

The energy band structure of common semiconductors near the center of the first Brillouin zone can be well described within the multiband EF approximation, which gives the following second-order Hamiltonian[8] of a Kramers-degenerate conduction band in the absence of external electric and magnetic fields

$$H_{k^2}^{(L)} = \sum_{q,q'} (-1)^{q+q'} k_{1-q} \vec{D}_{qq'} k_{1-q'} . \qquad (2)$$

Here $q, q' = \pm 1, 0$ and $k_{10} = k_z$, $k_{1\pm1} = \mp(k_x \pm ik_y)/\sqrt{2}$ represent cyclic components of the crystal momentum $\vec{k}$, and the superscript $(L)$ denotes the laboratory $L$-frame. The dyadic operator $\vec{D}_{qq'}$ acting on the EF-spinors (*slow variables*) is defined by its matrix elements in the basis of band-edge Bloch functions (*fast variables*)

$$<n'|\vec{D}_{qq'}|n> = (-1)^q \frac{\hbar^2}{2m_e} \delta_{n'n} \delta_{q-q'} - \frac{\hbar^2}{m_e^2} \sum_{\tilde{n}} \frac{<n'|p_{1q}|\tilde{n}><\tilde{n}|p_{1q'}|n>}{\varepsilon_{\tilde{n}} - \varepsilon_n}, \qquad (3)$$

where $\varepsilon_n$ is the energy at the bottom of the $n$-th band and $m_e$ is the bare electron mass. In the presence of a $k$-independent *intrinsic* SO coupling the Bloch functions $|n>$ are not factorizable into the orbital and spin parts, hence, the total angular momentum, $\vec{J} = \vec{L} + \vec{S}$, is required to characterize the basis kets. Within the "spherical approximation"[9], it is convenient to build the corresponding basis from the spherical spinor functions of the compound $L$-$S$ system[10] $|n> \equiv |LS, Jm; \vec{k} = 0> = \sum_{\mu,\mu_1} C^{Jm}_{L\mu\,1/2\,\mu_1} |L\mu>|1/2\,\mu_1>$, where $C^{Jm}_{L\mu\,1/2\,\mu_1}$ is the Clebsch-Gordan coefficient. The matrix elements of the direct tensor product $p_{1q} \cdot p_{1q'}$ in this basis are well known[10], which allows to calculate the second term in Eq.(3). Within Kane's eight-band model, which takes into account only the coupling of the conduction band ($L = 0, J = 1/2$) to the valence bands ($L = 1, J = 3/2$ and $L = 1, J = 1/2$), after some straightforward algebra, one finds

$$<L=0, S=1/2, J=1/2\, m_1 | H_{k^2}^{(L)} | L=0, S=1/2, J=1/2\, m> = \frac{\hbar^2 k^2}{2m_e}\delta_{mm_1} +$$

$$iP^2 \sum_{KQqq'}(-1)^K (K+1/2)^{1/2} C_{1/2\,m\,KQ}^{1/2\,m_1} C_{1q\,1q'}^{KQ} k_{1q} k_{1q'} \sum_{J_1 = 1/2, 3/2}(-1)^{3J_1}(2J_1+1)\begin{Bmatrix}1/2 & 1/2 & K \\ 1 & 1 & J_1\end{Bmatrix}\varepsilon_{J_1}^{-1}$$

(4)

Here $P = i\hbar <L=0 \| p_1 \| L=1>/(\sqrt{3}m_e)$ is the *reduced* Kane matrix element, which describes the coupling of the conduction and valence bands, and we set the energy of a conduction band to zero. Due to the triangle condition for the arguments $\{1/2, 1/2, K\}$ of the 6-$j$ symbol and Clebsch-Gordan coefficients, only scalar ($K = 0$) and vector ($K = 1$) terms are present in Eq.(4). The sum of the *scalar* terms on the RHS of Eq.(4) yields the Kane's expression for the reciprocal effective mass of a conduction electron

$$1/m^* = 1/m_e + (2/3\hbar^2)P^2[2E_g^{-1} + (E_g + \Delta)^{-1}], \qquad (5)$$

whereas the *vector* term gives $-iP^2[E_g^{-1} - (E_g + \Delta)^{-1}]\sum_Q (-1)^Q [\vec{k}\times\vec{k}]_{1-Q} C_{1/2\,m\,1Q}^{1/2\,m'}/\sqrt{3}$.

Here $E_g$ is the band-gap energy and $\Delta$ is the splitting of the valence band determined by the intrinsic SOC. Thus, Eq.(4) yields the following effective-mass Hamiltonian of the conduction band electron

$$H_{eff}^{(L)} = \frac{\hbar^2 k^2}{2m^*}\hat{I} - iP^2 \frac{\Delta}{3E_g(E_g+\Delta)}[\vec{k}\times\vec{k}]\cdot\vec{\sigma}, \qquad (6)$$

where $\hat{I}$ is the 2 x 2 unit matrix, and $\vec{\sigma}$ is acting on the spinor components of the conduction-band EF.

The Hamiltonian Eq.(6) represents the model where the crystal electron is treated as free spin-bearing particle moving in an effective magnetic field $\vec{B}_{eff} \sim \vec{k} \times \dot{\vec{k}}$. This field is zero for pure translational motion of a conduction electron and may not vanish only if its wave-vector is changing the direction in space. Suppose that the crystal momentum of a particle is rotating. The infinitesimal change in the direction of $\vec{k}$ can be described geometrically as $\delta \vec{k}^{(L)} = \delta\phi\,[\hat{n}^{(L)} \times \vec{k}^{(L)}]$, where $\delta\phi$ is the angle between $\vec{k}$ and $\vec{k} + \delta\vec{k}$ and $\hat{n}^{(L)}$ is the unit vector along the instantaneous axis of $\vec{k}$-rotation at instance $t$. In what follows, we choose the axis $\hat{n}^{(L)}$ to be at the right angle to the plane of the $\vec{k}$-rotation. Then $\delta\phi\,\hat{n}^{(L)} = \vec{k}^{(L)} \times \delta\vec{k}^{(L)} / k^2$ and one may define the *local instantaneous angular velocity* of this rotation as $\vec{\omega}_k^{(L)} = (d\phi/dt)\,\hat{n}^{(L)} = \vec{k}^{(L)} \times \dot{\vec{k}}^{(L)} / k^2$. We would like to emphasize here that this expression is purely kinematical, i.e., is independent of the dynamical cause of the $\vec{k}$-rotation.

The time dependence of $\vec{k}$ and, hence, the *L*-frame Hamiltonian Eq.(6), complicates the mathematical analysis of the problem. On the other hand, it is physically clear that in the reference frame that follows the rotation of $\vec{k}$ the Hamiltonian of the system will be time-independent. The unitary transformation $\Psi^{(R)}(t) = R^{(L)}(t)\,\Psi^{(L)}(t)$, where $\Psi$ is the instantaneous EF spinor, into the *R*-frame carried along by the infinitesimal rotation of $\vec{k}$ at the point $\vec{r}$ and time $t$ yields the following effective Hamiltonian $\widetilde{H}_{eff}^{(R)} = H_{eff}^{(R)} - \hbar \vec{\omega}_k \cdot \vec{j}$, where $H_{eff}^{(R)} = R H_{eff}^{(L)} R^{-1}$. We recall now that in the *R*-frame the Coriolis vector potential couples to the kinetic momentum of a particle, $\hbar \vec{k}^{(R)} = -i\hbar \vec{\nabla}^{(R)} - 2 m_e \vec{A}_{\omega_k}$, and it is easy to see that commutators of its components do not vanish, $[\vec{k} \times \vec{k}]^{(R)} = (2im_e/\hbar)\vec{\omega}_k^{(R)}$. Then it follows from Eq.(6) that the *R*-frame effective-mass Hamiltonian of the conduction-band electron can be expressed as (to first order in $\omega_k$ [11])

$$\tilde{H}_{eff}^{(R)} = \frac{\hbar^2 k^2}{2m^*}\hat{I} - \hbar\vec{\omega}_k \cdot \vec{S} + H_{SOR} \quad (7)$$

$$H_{SOR} = -\hbar \Delta g\, \vec{\omega}_k \cdot \vec{S}, \quad (8)$$

$$\Delta g = -4P^2 m_e \Delta/[3\hbar^2 E_g(E_g + \Delta)]. \quad (9)$$

Remarkably, the expression for $\Delta g$, Eq.(9), coincides with the Roth formula[12] for the deviation of the *g*-factor of a conduction electron in bulk semiconductors from its free value. The Hamiltonian Eq.(7) depends on the choice of guide that specifies the reference orientation, i.e. the orientation in which the *R*-frame coincides with some space-fixed frame. At the moment $t = 0$, the reference orientation may always be chosen such that $z_M$ coincides with $z_L$, which determines our gauge convention. Note that if the rotation is uniform, the operator $R = \exp(i\vec{\omega}_k \vec{j}t)$ maps the *L*-frame into the actual orientation of the *R*-frame at any time *t*.

Comparison of Eq.(7) with Eq.(1) shows that the account of SO coupling in the system yields in addition to Mashhoon spin-rotation interaction in the reciprocal momentum space, $-\hbar\vec{\omega}_k \vec{S}$, the term $H_{SOR}$, Eq.(8). It corresponds to the SOR-coupling in the *k*-space and is the same from the point of view of a rotating as well as inertial observers. Indeed, to describe the evolution of the EF in the *local* inertial frame we have to perform a reverse rotation of the coordinate system compensating for the rotation of the *R*-frame. This transformation is not associated with a physical change of a state and does not affect the isotropic kinetic energy of the carrier. At any moment at time, it is merely the operator of a reverse rotation in the 2x2 spinor-space, which yields

$$H_{eff}^{(L)} = \frac{\hbar^2 k^2}{2m^*}\hat{I} + H_{SOR} \qquad (10)$$

The SOR Hamiltonian Eq.(8) represents the weak-SO-limit of the effective Hamiltonian obtained in Ref.[6] by a more general method. Notably, $H_{SOR}$ is analogous to the usual spin-rotation interaction in molecular systems. The analogy becomes exact if one replace $\omega_k$ with the angular velocity of a molecular frame in the real space. Formally, the same spin-Hamiltonian governs the evolution of the Kramers-doublet in an adiabatically *revolving* external electric field in the limit of a weak SO interaction in the valence bands[13]. Fundamentally, these very diverse physical phenomena can be described uniformly in purely geometric terms as a consequence of the difference in the Berry phase acquired by the components of the spin-orbitally mixed Kramers-doublet during its cyclic evolution in the relevant parameter space. The geometric interpretation of the SOR interaction in molecular systems has been given in Ref.[14] and was extended to semiconductors in Refs.[6, 13]. The Berry phase effects in semiconductors emerging from the SO coupling have been proposed[15] to occur as early as in 1993. Now it is well recognized that an adiabatic change in the direction of the wave vector of a charge carrier leads to non-trivial gauge potentials that appear in the reciprocal momentum space. The associated covariant gauge field enters the equation of motion for the group velocity of a wave-packet and may affect the coherent transport properties of charge carriers[16,17,18,19,20,21]. However, until recently no connection was made between the results of these studies and manifestation of SOR interaction in semiconductors.

The straightforward derivation presented here strengthens the foundation of the analysis made in Ref.[6], which we would like to briefly summarize here. During the "slow" adiabatic ($\hbar|\vec{\omega}_k| \ll |E_g|$) rotation of the envelope wave function the spin of a conduction-band electron will follow the rotation of the EF wave-vector with some slippage determined by $\Delta g$. In the *L*-frame, this process can be described as a spin precession in an effective magnetic field $\vec{B}_{eff} = -(\hbar/\mu_B)\Delta g \vec{\omega}_k = -(\hbar/\mu_B)\Delta g(\vec{k} \times \dot{\vec{k}}/k^2)$. This field is orthogonal to the plane of particle's rotation, is related neither to Rashba nor to Dresselhaus couplings between the spin of the charge carrier and its momentum, and may appear in spherically symmetric bulk crystals. *Dynamic* anisotropy of a system *locally* in the *k*-space ($P \neq 0$) is the fundamental precondition for manifestation of this interaction in the conduction band. Although the SOR interaction in the conduction band appears already in the second-order, the small prefactor, $\hbar/\mu_B \cong 10^{-7} G \cdot \sec$, makes direct gyromagnetic experiment of Barnett- or mechanical Faraday-type[22] a rather challenging task. On the other hand, an electron can be forced to rotate rapidly by external and/or internal electric fields. Whereas the latter, e.g., through collisions with crystal impurities will lead to Elliott spin dephasing[23][24], the former may be used to control the spin splitting and precession at *B* = 0. It is easy to see that if the uniform electric field is the sole source of the electron's acceleration perpendicular to its instantaneous velocity, then $H_{SOR} = \Delta g(\vec{k} \times e\vec{E}/k^2)\vec{S}$ and the magnitude of the SOR coupling can be comparable or larger than Rashba or Dresselhaus interactions[7], e.g., for $k = 10^8 m^{-1}$ and $E = 10^5 V/m$, $H_{SOR} = \Delta g\ meV$.